\documentstyle[12pt]{article}
\def\be {\begin{equation}}
\def\ee {\end{equation}}
\def\ba {\begin{eqnarray}}
\def\ea {\end{eqnarray}}
\newcommand{\bq}{\begin{eqnarray}}
\newcommand{\eq}{\end{eqnarray}}

\def\bi {\begin{itemize}}
\def\ei {\end{itemize}}
\begin{document}
\def\bea{\begin{eqnarray}}
\def\eea{\end{eqnarray}}
\title{\bf  {Interacting  holographic dark energy model and generalized second law of thermodynamics
 in non-flat universe}}
 \author{M.R. Setare  \footnote{E-mail: rezakord@ipm.ir}
  \\{Department of Science,  Payame Noor University. Bijar. Iran}}
\date{\small{}}

\maketitle
\begin{abstract} In the present paper we consider
the interacting holographic model of dark energy to investigate the
validity of the generalized second laws of thermodynamics in
non-flat (closed) universe enclosed by  the event horizon measured
from the
 sphere of the horizon named $L$. We show that for $L$ as the system's IR cut-off  the generalized
 second law is respected for the special range of the deceleration
parameter.
 \end{abstract}

\newpage

\section{Introduction}
One of the most important problems of cosmology, is the problem of
so-called dark energy (DE). The type Ia supernova observations
suggests that the universe is dominated by dark energy with negative
pressure which provides the dynamical mechanism of the accelerating
expansion of the universe \cite{{per},{gar},{ries}}. The strength of
this acceleration is presently matter of debate, mainly because it
depends on the theoretical model implied when interpreting the data.
Most of these models are based on dynamics of a scalar or
multi-scalar fields (e.g quintessence \cite{{rat},{zlat}} and
quintom model of dark energy,
respectively). \\
An approach to the problem of DE arises from the holographic
principle that states that the number of degrees of freedom related
directly to entropy scales with the enclosing area of the system. It
was shown by 'tHooft and Susskind \cite{hologram} that effective
local quantum field theories greatly overcount degrees of freedom
because the entropy scales extensively for an effective quantum
field theory in a box of size $L$ with UV cut-off $ \Lambda$. As
pointed out by \cite{myung}, attempting to solve this problem, Cohen
{\it et al} showed \cite{cohen} that in quantum field theory, short
distance cut-off $\Lambda$ is related to long distance cut-off $L$
due to the limit set by forming a black hole. In other words the
total energy of the system with size $L$ should not exceed the mass
of the same size black hole, i.e. $L^3 \rho_{\Lambda}\leq LM_p^2$
where $\rho_{\Lambda}$ is the quantum zero-point energy density
caused by UV cut-off $\Lambda$ and $M_P$ denotes the Planck mass (
$M_p^2=1/{8\pi G})$. The largest $L$ is required to saturate this
inequality. Then its holographic energy density is given by
$\rho_{\Lambda}= 3c^2M_p^2/8\pi L^2$ in which $c$ is a free
dimensionless parameter and coefficient 3 is for convenience.

 As an application of the holographic principle in cosmology,
 it was studied by \cite{KSM} that the consequence of excluding those degrees of freedom of the system
 which will never be observed by the effective field
 theory gives rise to IR cut-off $L$ at the
 future event horizon. Thus in a universe dominated by DE, the
 future event horizon will tend to a constant of the order $H^{-1}_0$, i.e. the present
 Hubble radius. The consequences of such a cut-off could be
 visible at the largest observable scales and particularly in the
 low CMB multipoles where we deal with discrete wavenumbers. Considering the power spectrum in a finite
 universe as a consequence of a holographic constraint, with different boundary
 conditions, and fitting it with LSS, CMB and supernova data, a cosmic duality between the dark energy equation of state
 and power spectrum is obtained that can describe the low $l$ features extremely
 well.

 On the basis of the cosmological state of the holographic principle, proposed by Fischler and
Susskind \cite{fischler}, a holographic model of dark Energy (HDE)
has been proposed and studied widely in the
 literature \cite{miao,HDE}. In \cite{HG} using the type Ia
 supernova data, the model of HDE is constrained once
 when c is unity and another time when c is taken as a free
 parameter. It is concluded that the HDE is consistent with recent observations, but future observations are needed to
 constrain this model more precisely. In another paper \cite{HL},
 the anthropic principle for HDE is discussed. It is found that,
 provided that the amplitudes of fluctuation are variable, the
 anthropic consideration favors the HDE over the cosmological
 constant.

 In HDE, in order to determine the proper and well-behaved system's IR cut-off, there are some
difficulties that must be studied carefully to get results adapted
with experiments that claim our universe has accelerated expansion.
For instance, in the model proposed by \cite{miao}, it is discussed
that considering  the particle horizon, as the IR cut-off, the HDE
density reads
 \be
  \rho_{\Lambda}\propto a^{-2(1+\frac{1}{c})},
\ee
 that implies $w>-1/3$ which does not lead to an accelerated
universe. Also it is shown in \cite{easther} that for the case of
closed
universe, it violates the holographic bound.\\
The problem of taking apparent horizon (Hubble horizon) - the
outermost surface defined by the null rays which instantaneously are
not expanding, $R_A=1/H$ - as the IR cut-off in the flat universe
was discussed by Hsu \cite{Hsu}. According to Hsu's argument,
employing the Friedmann equation $\rho=3M^2_PH^2$ where $\rho$ is
the total energy density and taking $L=H^{-1}$ we will find
$\rho_m=3(1-c^2)M^2_PH^2$. Thus either $\rho_m$ or $\rho_{\Lambda}$
behave as $H^2$. So the DE results as pressureless, since
$\rho_{\Lambda}$ scales like matter energy density $\rho_m$ with the
scale factor $a$ as $a^{-3}$. Also, taking the apparent horizon as
the IR cut-off may result in a constant parameter of state $w$,
which is in contradiction with recent observations implying variable
$w$ \cite{varw}.
 On the other hand taking the event horizon, as
 the IR cut-off, gives results compatible with observations for a flat
 universe.

 It is fair to claim that the simplicity and reasonable nature of HDE
 provide a
 more reliable framework  for investigating the problem of DE compared with other models
proposed in the literature\cite{cosmo,quint,phant}. For instance the
coincidence or "why now?" problem is easily solved in some models of
HDE based on this fundamental assumption that matter and holographic
dark energy do not conserve separately, but the matter energy
density
decays into the holographic energy density \cite{interac}.\\
Some experimental data have implied that our universe is not a
perfectly flat universe and recent papers have favored a universe
with spatial curvature \cite{curve}. As a matter of fact, we want to
remark that although it is believed that our universe is flat, a
contribution to the Friedmann equation from spatial curvature is
still possible if the number of e-foldings is not very large
\cite{miao2}. Defining the appropriate distance for the case of a
non-flat universe is another story. Some aspects of the problem have
been discussed in \cite{miao2,guberina}. In this case, the event
horizon can not be considered as the system's IR cut-off, because,
for instance, when the dark energy is dominated and $c=1$, where $c$
is a positive constant, $\Omega_\Lambda=1+ \Omega_k$, we find $\dot
R_h<0$, while we know that in this situation we must be in de Sitter
space with a constant EoS. To solve this problem, another distance
is considered- the radial size of the event horizon measured on the
sphere of the horizon, denoted by $L$- and the evolution of the
holographic model of dark energy in a non-flat universe is
investigated.
\\
Since the discovery of black hole thermodynamics in 1970 physicists
have speculated on the thermodynamics of cosmological models in an
accelerated expanding universe \cite{thermo}. Related to the present
work, in \cite{abdalla}, for time independent and timedependent
equations of state (EoS), the first and second laws of
thermodynamics in a flat universe were investigated. For the case of
a constant EoS, the first law is valid for the apparent horizon
(Hubble horizon) and it does not hold for the event horizon as
system's IR cut-off. When the EoS is assumed to be time dependent,
using a holographic model of dark energy in flat space, the same
result is gained: the event horizon, in contrast to the apparent
horizon, does not satisfy the first law. Also, while the event
horizon does not respect the second law, it hold for the universe enclosed by the apparent horizon.\\
In previous paper \cite{ss} we have investigated the validity of the
first and second laws of thermodynamic in a non-flat universe
enclosed by the apparent horizon $R_A$ and the event horizon
measured from the sphere of the horizon named $L$. In the present
paper we extend this investigation to the interacting holographic
model of dark energy in a non-flat universe, we study the validity
of generalized second law (GSL) of thermodynamics in present time
for a universe enveloped by $L$.
\section{ Intracting holographic dark energy density }
In this section we obtain the equation of state for the holographic
energy density when there is an interaction between holographic
energy density $\rho_{\Lambda}$ and a Cold Dark Matter(CDM) with
$w_{m}=0$. The continuity equations for dark energy and CDM are
\begin{eqnarray}
\label{2eq1}&& \dot{\rho}_{\rm \Lambda}+3H(1+w_{\rm \Lambda})\rho_{\rm \Lambda} =-Q, \\
\label{2eq2}&& \dot{\rho}_{\rm m}+3H\rho_{\rm m}=Q.
\end{eqnarray}
The interaction is given by the quantity $Q=\Gamma
\rho_{\Lambda}$. This is a decaying of the holographic energy
component into CDM with the decay rate $\Gamma$. Taking a ratio
of two energy densities as $r=\rho_{\rm m}/\rho_{\rm \Lambda}$,
the above equations lead to
\begin{equation}
\label{2eq3} \dot{r}=3Hr\Big[w_{\rm \Lambda}+
\frac{1+r}{r}\frac{\Gamma}{3H}\Big]
\end{equation}
 Following Ref.\cite{Kim:2005at},
if we define
\begin{eqnarray}\label{eff}
w_\Lambda ^{\rm eff}=w_\Lambda+{{\Gamma}\over {3H}}\;, \qquad w_m
^{\rm eff}=-{1\over r}{{\Gamma}\over {3H}}\;.
\end{eqnarray}
Then, the continuity equations can be written in their standard
form
\begin{equation}
\dot{\rho}_\Lambda + 3H(1+w_\Lambda^{\rm eff})\rho_\Lambda =
0\;,\label{definew1}
\end{equation}
\begin{equation}
\dot{\rho}_m + 3H(1+w_m^{\rm eff})\rho_m = 0\; \label{definew2}
\end{equation}
We consider the non-flat Friedmann-Robertson-Walker universe with
line element
 \be\label{metr}
ds^{2}=-dt^{2}+a^{2}(t)(\frac{dr^2}{1-kr^2}+r^2d\Omega^{2}).
 \ee
where $k$ denotes the curvature of space k=0,1,-1 for flat, closed
and open universe respectively. A closed universe with a small
positive curvature ($\Omega_k\sim 0.01$) is compatible with
observations \cite{curve}. We use the Friedmann equation to relate
the curvature of the universe to the energy density. The first
Friedmann equation is given by
\begin{equation}
\label{2eq7} H^2+\frac{kc^2}{a^2}=\frac{1}{3M^2_p}\Big[
 \rho_{\rm \Lambda}+\rho_{\rm m}\Big].
\end{equation}
Define as usual
\begin{equation} \label{2eq9} \Omega_{\rm
m}=\frac{\rho_{m}}{\rho_{cr}}=\frac{ \rho_{\rm
m}}{3M_p^2H^2},\hspace{1cm}\Omega_{\rm
\Lambda}=\frac{\rho_{\Lambda}}{\rho_{cr}}=\frac{ \rho_{\rm
\Lambda}}{3M^2_pH^2},\hspace{1cm}\Omega_{k}=\frac{kc^2}{a^2H^2}
\end{equation}
Now we can rewrite the first Friedmann equation as
\begin{equation} \label{2eq10} \Omega_{\rm m}+\Omega_{\rm
\Lambda}=1+\Omega_{k}.
\end{equation}
Using Eqs.(\ref{2eq9},\ref{2eq10}) we obtain following relation
for ratio of energy densities $r$ as
\begin{equation}\label{ratio}
r=\frac{1+\Omega_{k}-\Omega_{\Lambda}}{\Omega_{\Lambda}}
\end{equation}
In non-flat universe, our choice for holographic dark energy
density is
 \be \label{holoda}
  \rho_\Lambda=3c^2M_{p}^{2}L^{-2}.
 \ee
As it was mentioned, $c$ is a positive constant in holographic model
of dark energy($c\geq1$) and the coefficient 3 is for convenient.
$L$ is defined as the following form:
\begin{equation}\label{leq}
 L=ar(t),
\end{equation}
here, $a$, is scale factor and $r(t)$ can be obtained from the
following equation
\begin{equation}\label{rdef}
\int_0^{r(t)}\frac{dr}{\sqrt{1-kr^2}}=\int_t^\infty
\frac{dt}{a}=\frac{R_h}{a},
\end{equation}
where $R_h$ is event horizon. Therefore while $R_h$ is the radial
size of the event horizon measured in the $r$ direction, $L$ is the
radius of the event horizon measured on the sphere of the
horizon.\footnote{In Reference \cite{zp}, the interacting
holographic dark energy model in non-flat FRW universes have been
studied. The authors of \cite{zp} have considered the Hubble length
as infrared cut-off, but as I have discussed in introduction:
employing the Friedmann equation $\rho=3M^2_PH^2$ where $\rho$ is
the total energy density and taking $L=H^{-1}$ we will find
$\rho_m=3(1-c^2)M^2_PH^2$. Thus either $\rho_m$ or $\rho_{\Lambda}$
behave as $H^2$. So the DE results as pressureless, since
$\rho_{\Lambda}$ scales like matter energy density $\rho_m$ with the
scale factor $a$ as $a^{-3}$. Also, taking the apparent horizon as
the IR cut-off may result in a constant parameter of state $w$,
which is in contradiction with recent observations implying variable
$w$ \cite{varw}.}
 For closed universe we have (same calculation is valid for
open universe by transformation)
 \be \label{req}
 r(t)=\frac{1}{\sqrt{k}} sin y.
 \ee
where $y\equiv \sqrt{k}R_h/a$. Using definitions
$\Omega_{\Lambda}=\frac{\rho_{\Lambda}}{\rho_{cr}}$ and
$\rho_{cr}=3M_{p}^{2}H^2$, we get

\begin{equation}\label{hl}
HL=\frac{c}{\sqrt{\Omega_{\Lambda}}}
\end{equation}
Now using Eqs.(\ref{leq}, \ref{rdef}, \ref{req}, \ref{hl}), we
obtain
 \be \label{ldot}
 \dot L= HL+ a \dot{r}(t)=\frac{c}{\sqrt{\Omega_\Lambda}}-cos y,
\end{equation}
By considering  the definition of holographic energy density
$\rho_{\rm \Lambda}$, and using Eqs.( \ref{hl}, \ref{ldot}) one
can find:
\begin{equation}\label{roeq}
\dot{\rho_{\Lambda}}=-2H(1-\frac{\sqrt{\Omega_\Lambda}}{c}\cos
y)\rho_{\Lambda}
\end{equation}
Substitute this relation into Eq.(\ref{2eq1}) and using
definition $Q=\Gamma \rho_{\Lambda}$, we obtain
\begin{equation}\label{stateq}
w_{\rm \Lambda}=-(\frac{1}{3}+\frac{2\sqrt{\Omega_{\rm
\Lambda}}}{3c}\cos y+\frac{\Gamma}{3H}).
\end{equation}
Here as in Ref.\cite{WGA}, we choose the following relation for
decay rate
\begin{equation}\label{decayeq}
\Gamma=3b^2(1+r)H
\end{equation}
with  the coupling constant $b^2$. Using Eq.(\ref{ratio}), the
above decay rate take following form
\begin{equation}\label{decayeq2}
\Gamma=3b^2H\frac{(1+\Omega_{k})}{\Omega_{\Lambda}}
\end{equation}
Substitute this relation into Eq.(\ref{stateq}), one finds the
holographic energy equation of state
\begin{equation} \label{3eq4}
w_{\rm \Lambda}=-\frac{1}{3}-\frac{2\sqrt{\Omega_{\rm
\Lambda}}}{3c}\cos y-\frac{b^2(1+\Omega_{k})}{\Omega_{\rm \Lambda}}.
\end{equation}
\section{Generalized Second Law of Thermodynamics}

Here, we study the validity of a GSL of thermodynamics. According to
the GSL, for our system, the sum of the entropy of matter enclosed
by the horizon and the entropy of the horizon must not be a
decreasing function of time. We investigate this law for the
universe filled with perfect fluid described by a normal scalar
(quintessence-like) field. For this purpose, we consider the
enclosed matter and calculate its entropy.

Before going into mathematics of the GSL, we want to work out
remarkable points as regard the temperature of the fluid. According
to the generalization the black hole thermodynamics to our
cosmological model, we have taken the temperature of our horizon to
be $T_L=(1/2\pi L)$ where $L$ denotes the size of the universe. In
investigating the GSL, the definition of the temperature of the
fluid needs further discussions. The only temperature to hand is the
horizon temperature. If the fluid temperature is equal to the
horizon temperature, the system will be in equilibrium. Another
possibility \cite{davies2} is that the fluid temperature is
proportional to the horizon temperature i.e. for the fluid enveloped
by the apparent horizon $T=eH/2\pi$ \cite{pavon}. In general, the
systems must interact for some length of time before they can attain
thermal equilibrium. In the case at hand, the interaction certainly
exists as any variation in the energy density and/or pressure of the
fluid will automatically induce a modification of the horizon radius
via Einstein's equations. Moreover if $T \neq T_{L}$, then energy
would spontaneously flow between the horizon and the fluid (or
viceversa), something at variance with the FRW geometry \cite{pa}.
In general, when we consider the thermal equilibrium state of the universe,
the temperature of the universe is associated with the horizon.\\
We consider the FRW universe as a thermodynamical system with
horizon surface as a boundary of the system. In general the radius
of the event horizon $L$ is not constant but changes with time. Let
$dL$ be an infinitesimal change of the event horizon radius during a
time of interval $dt$. This small displacement $dL$ will cause a
small change $dV$ in the volume $V$ of the event horizon. This leads
to build up two systems of space-time with radii $L$ and $L + dL $
having a common source $T_{\mu\nu}$ of perfect fluid with non-zero
pressure $P$ and energy density $\rho$. Each space-time describing a
thermodynamical system and satisfying Einstein equations, differs
infinitesimally in the extensive variables volume, energy and
entropy by $dV$, $dE$ and $dS$, respectively, while having same
values the intensive variables temperature $T$ and pressure $P$.
Thus, for these two space-times describing two thermodynamical
states, there must exist some relation among these thermodynamic
quantities. It turns out it is indeed the case: the differential
form of the Friedman equation can be rewritten to a universal form,
$dE=TdS +PdV$. Therefore the entropy of the holographic energy  and
CDM have the following relations respectively with them pressure and
energy \cite{{abdalla},{gww}}
 \be \label{en1}
 dS_{\Lambda}=\frac{1}{T}\ (P_{\Lambda}dV+dE_{\Lambda})
 \ee
 \be \label{en2}
 dS_{m}=\frac{1}{T}\ (P_{m}dV+dE_{m})
 \ee
 where $V=4 \pi L^3/3$ is the volume containing the matter and dark
 energy.
 \be \label{ene1}
 E_{\Lambda}=4 \pi L^3 \rho_{\Lambda}/3, \hspace{1cm} P_{\Lambda}=w_\Lambda^{\rm eff}\rho_{\Lambda}
  \ee
\be \label{ene2}
 E_{m}=4 \pi L^3 \rho_{m}/3, \hspace{1cm} P_{m}=w_{m}^{\rm eff}\rho_{m}
  \ee
Now using Eqs.(\ref{eff}, \ref{2eq9} ,\ref{holoda})we have
\footnote{Using $T_L=(1/2\pi L)$, we get \be dS_{\Lambda}= 24
\pi^{2} c^2 M_{P}^{2}(1+w_\Lambda)L dL \ee Thus we see that the
entropy enveloped by the event horizon is $S_{\Lambda}\sim L^2$.}
\be E_{\Lambda}=4\pi c^2 M_{P}^{2}L, \hspace{1cm} P_{\Lambda}=3c^2
M_{P}^{2}L^{-2} (w_{\Lambda}+\frac{\Gamma}{3H}) \label{ep1} \ee

\be E_{m}=4\pi c^2 M_{P}^{2}L\frac{\Omega_{m}}{\Omega_{\Lambda}},
\hspace{1cm} P_{m}=\frac{-\Gamma \Omega_{m}M_{P}^{2}H}{r}
\label{ep2} \ee
 Taking
derivative in both sides of (\ref{en1}) with respect to $x(\equiv
lna)$, using Eqs.(\ref{hl}, \ref{ldot}, \ref{stateq}) and after some
calculation, we obtain \be \label{ende1}
\frac{dS_{\Lambda}}{dx}=\frac{16\pi^{2}c^3 M_{P}^{2}}{H^2
\sqrt{\Omega_{\Lambda}}}(\frac{\sqrt{\Omega_{\Lambda}}}{C}\cos^{2}y-\cos
y) \ee Now we taking derivative in both sides of (\ref{en2}) with
respect to $x(\equiv lna)$, after some calculation, we obtain
 \be \label{ende2}\frac{dS_{m}}{dx}=8\pi^{2}M_{P}^{2}L[-L^2\dot{L}\frac{\Gamma
 \Omega_{m}}{r}+c^2[\frac{ \Omega_{m}\dot{L}}{\Omega_{\Lambda}H}+\frac{L\dot{\Omega_{m}}}{\Omega_{\Lambda}H}
 -\frac{L \Omega_{m}}{H\Omega_{\Lambda}^{2}}\dot{\Omega_{\Lambda}}]]
\ee Now we go to obtain $\dot{\Omega_{m}}$, taking derivative in
both sides of (\ref{2eq10}), we obtain \be \label{dotk}
\dot{\Omega_{m}}=\dot{\Omega_{k}}-\dot{\Omega_{\Lambda}}  \ee Using
Eq.(\ref{ratio}), one can obtain following relation for
$\dot{\Omega_{k}}$ \be \label{omegah}
\dot{\Omega_{k}}=\dot{r}\Omega_{\Lambda}+\frac{\dot{\Omega_{\Lambda}}}{\Omega_{\Lambda}}+\frac{\dot{\Omega_{\Lambda}}
\Omega_{k}}{\Omega_{\Lambda}} \ee By use this equation and Eq.(\ref
{2eq3}), we obtain \be \label{dotm}
\dot{\Omega_{m}}=\dot{\Omega_{\Lambda}}(1-\frac{1}{\Omega_{\Lambda}}-\frac{\Omega_{k}}{\Omega_{\Lambda}})+
3Hr\Omega_{\Lambda}\Big[w_{\rm \Lambda}+
\frac{1+r}{r}\frac{\Gamma}{3H}\Big] \ee Also, using Eqs.(\ref{hl},
\ref{ldot}), we have \be \label{DOTOML}
\dot{\Omega_{\Lambda}}=-2\Omega_{\Lambda}[(1-\frac{\sqrt{\Omega_{\Lambda}}}{c}\cos
y)H+\frac{\dot{H}}{H}] \ee Using the above relations, we find
following expression

\bq \label{ende3}\frac{dS_{m}}{dx}=\nonumber
\frac{8\pi^{2}M_{P}^{2}c^3}{H^2\Omega_{\Lambda}^{3/2}}[[-3b^2
(1+\Omega_{k})+(1+\Omega_{k}-\Omega_{\Lambda})](\frac{c}{\sqrt{\Omega_\Lambda}}-cos
y)\\+\frac{c}{H\sqrt{\Omega_{\Lambda}}}[4(1+\Omega_{k}-\Omega_{\Lambda})[(1-\frac{\Omega_{\Lambda}}{c}\cos
y)H+\frac{\dot{H}}{H}]+3\Omega_{\Lambda}Hr(w_{\Lambda}+b^2\frac{(1+r)^{2}}{r})]
\eq We can rewrite the above equation in terms of deceleration
parameters $q$ as \bq \label{ende4}\frac{dS_{m}}{dx}=\nonumber
\frac{8\pi^{2}M_{P}^{2}c^3}{H^2\Omega_{\Lambda}^{3/2}}[[-3b^2
(1+\Omega_{k})+(1+\Omega_{k}-\Omega_{\Lambda})](\frac{c}{\sqrt{\Omega_\Lambda}}-cos
y)\\+\frac{c}{H^2\sqrt{\Omega_{\Lambda}}}[4(1+\Omega_{k}-\Omega_{\Lambda})[(1-\frac{\Omega_{\Lambda}}{c}\cos
y)-(1+q)]+(1+\frac{2\sqrt{\Omega_{\Lambda}}}{c}\cos
y+b^2(1+\Omega_{k})]] \eq where \be\label{q}
 q=-\frac{\dot H}{H^2}-1 \ee
 The entropy of horizon, is
$S_L=\pi L^2$, so one can easily find
\be\label{dSA}\frac{dS_{L}}{dx}=\frac{2\pi
c}{\sqrt{\Omega_{\Lambda}}H^2}(\frac{c}{\sqrt{\Omega_\Lambda}}-cosy)
\ee
 From the equations (\ref{ende1} , \ref{ende4}, \ref{dSA}) it is obtained
 \bq\label{dSTOTA}
 \frac{d}{dx}(S_{\Lambda}+S_{m}+S_{L})=\nonumber\frac{16\pi^{2}c^3 M_{P}^{2}}{H^2
\sqrt{\Omega_{\Lambda}}}(\frac{\sqrt{\Omega_{\Lambda}}}{C}\cos^{2}y-\cos
y)+ \frac{8\pi^{2}M_{P}^{2}c^3}{H^2\Omega_{\Lambda}^{3/2}}[[-3b^2
(1+\Omega_{k})\\+\nonumber(1+\Omega_{k}-\Omega_{\Lambda})](\frac{c}{\sqrt{\Omega_\Lambda}}-cos
y)+\frac{c}{H^2\sqrt{\Omega_{\Lambda}}}[4(1+\Omega_{k}-\Omega_{\Lambda})[(1-\frac{\Omega_{\Lambda}}{c}\cos
y)-(1+q)]\\+(1+\frac{2\sqrt{\Omega_{\Lambda}}}{c}\cos
y+b^2(1+\Omega_{k}))]+\frac{2\pi
c}{\sqrt{\Omega_{\Lambda}}H^2}(\frac{c}{\sqrt{\Omega_\Lambda}}-cosy)
 \eq
 We restrict our consideration to present time, $\Omega_\Lambda=
0.73$, $\Omega_{k}=0.01$ and take $c$ to be 1. According to
Eqs.(\ref{2eq9}, \ref{holoda}, \ref{leq}, \ref{req}), the ratio of
the energy density between curvature and holographic dark energy is
\be \label{sin}
\frac{\Omega_{k}}{\Omega_{\Lambda}}=\frac{\rho_{k}}{\rho_{\Lambda}}=\frac{kL^2}{a^2}=kr^2=\sin^{2}
y \ee Then \be \label{cos} \cos
y=\sqrt{1-\frac{\Omega_{k}}{\Omega_{\Lambda}}} \ee therefore in the
present time $\cos y=0.99$.  To determine the sign of $q$, we pay
attention that amounts that $q$ can take, depend on the sign of
$\dot H$. For $\dot H>0$ we have phantom-like behavior and from
previous discussions we know since the Hawking-Gibbon's bound does
not allow our holographic DE model to be of this kind, we have to
rule out this possibility. If $\dot H=0$ we are in de- Sitter
space-time and $q=-1$, therefore $\frac{d}{dx}(S_{\Lambda}+
S_m+S_L)=0$. Eventually for the case of $\dot H<0$ we find $q>-1$.
 Simplifying
(\ref{dSTOTA}) by means of putting values of $\Omega_\Lambda=0.73$,
$\Omega_{k}=0.01$, $b^{2}=0.2$ and $c=1$  makes the following form
 \be\label{ESTdS}
 \frac{d}{dx}(S_{\Lambda}+
S_m+S_L)=\frac{M_{P}^{2}}{H^2}(-10.88+\frac{1482.88-167.42
q}{H^2})+\frac{1.33}{H^2}.
 \ee
 From (\ref{ESTdS}) we find that for $q\leq 8.85-\frac{H^2}{15.4}$ then $\frac{d}{dx}(S_{\Lambda}+
S_m+S_L)>0$.
 \section{Conclusions}
In 1973, Bekenstein \cite{bek} assumed that there is a relation
between the event of horizon and the thermodynamics of a black hole,
so that the event of horizon of the black hole is a measure of the
entropy of it. Along this line of thought, it was argued in
\cite{jac} that the gravitational Einstein equations can be derived
through a thermodynamical argument using the relation between area
and entropy as input. Following \cite{{jac},{fro}}, Danielsson
\cite{dan}has been able to obtain the Friedmann equations, by
applying the relation $\delta Q=T dS$ to a cosmological horizon to
calculate the heat flow through the horizon in an expanding universe
in an acceleration phase.
 This idea has been generalized to horizons of
cosmological models, so that each horizon corresponds to an entropy.
Thus the second law of thermodynamics was modified in the way that
in generalized form, the sum of all time derivative of entropies
related to horizons plus time derivative of normal entropy must be
positive i.e. the sum of entropies must be increasing function of
time. In \cite{davies2}, the validity of generalized second law for
the cosmological models which departs slightly from de Sitter space
is investigated. However, it is only natural to associate an entropy
to the horizon area as it measures our lack of knowledge about what
is going on beyond it. It is of interest to remark that in the
literature, the different scenarios of DE has never been studied via
considering special similar horizon, as in \cite{davies2} the
apparent horizon, $1/H$, determines our universe while in
\cite{gong}, in the Brans-Dicke cosmology framework, the universe is
enclosed by event horizon, $R_h$. Thus it looks that we need to
define a horizon that satisfies all of our accepted principles; in
\cite{odintsov} a linear combination of event and apparent horizon,
as IR cut-off has been considered.

In present paper, we studied $L$, as the horizon measured from the
sphere of the horizon as system's IR cut-off.  We investigated the
GSL of thermodynamics at present time for the universe enveloped by
this horizon. In our study we assumed that the universe is in
thermal equilibrium, also we considered an interaction between
holographic energy density and CDM. We obtained that the GSL was
satisfied just for a range of $q$ which $q$ was deceleration
parameter.

\end{document}